\documentclass[%
 reprint,
 amsmath,amssymb,
 prl
]{revtex4-2}

\usepackage[paperwidth=210mm,paperheight=297mm,centering,hmargin=2cm,vmargin=2.5cm]{geometry}
\usepackage{dcolumn}% Align table columns on decimal point
\usepackage{bm}% bold math
\usepackage{hyperref}% add hypertext capabilities
\usepackage[mathlines]{lineno}% Enable numbering of text and display math
\usepackage{amsmath}
\usepackage{amssymb}
\usepackage{graphicx,bbm,mathrsfs}
\usepackage{nicefrac}
\usepackage{slashed}
\usepackage{bbm}

\usepackage{empheq}
\usepackage{tikz}
\usepackage{tcolorbox}

\begin{document}

\title{The LHC as an Axion-Photon Collider}% Force line breaks with \\

\author{Sergio Barbosa}
 \email{sergio.barbosa@aluno.ufabc.edu.br}
\affiliation{CCNH, Universidade Federal do ABC, Santo Andr\'e, 09210-580 SP, Brazil}

\author{Matheus Coelho}%
 \email{matheus.pereira.coelho@cern.ch}
\affiliation{CCNH, Universidade Federal do ABC, Santo Andr\'e, 09210-580 SP, Brazil}%

\author{Sylvain Fichet}
 \email{sylvain.fichet@gmail.com}
\affiliation{CCNH, Universidade Federal do ABC, Santo Andr\'e, 09210-580 SP, Brazil}%

\author{Gustavo Gil da Silveira}
 \email{gustavo.silveira@cern.ch}
\affiliation{CERN, PH Department, 1211 Geneva, Switzerland}
\affiliation{Experimental Group of the CMS Collaboration (GECMS), Caixa Postal 15051, Porto Alegre, RS CEP 91501-970, Brazil}

\author{Magno Machado}
 \email{magno.machado@ufrgs.br}
\affiliation{High Energy Physics Phenomenology Group, GFPAE IF-UFRGS, Caixa Postal 15051, Porto Alegre, RS CEP 91501-970, Brazil}

\begin{abstract}

Assuming the existence of an axion-like particle (ALP), beams of relativistic particles emit fluxes of quasi-real ALPs, analogous to the photon fluxes  described by Weizs\"acker-Williams-type approximations.  Consequently,
ALP-ALP and ALP-photon collisions can occur at the LHC. 
We initiate the study of the LHC as an ALP collider, and show that ALP collisions provide competitive probes of certain ALP couplings. We show that ALP fluxes from heavy ions are suppressed relative to those from protons, unlike their photon counterpart. As a result, the most likely processes are ALP-photon collisions occurring in the proton-ion ($p$A) ultraperipheral collisions. Using our implementation of ALP fluxes in simulation tools, we show that ALP-photon collisions in $p$Pb efficiently probe  ALP couplings to third generation fermions. 
LHC data with realistic  $p$Pb luminosity  can constrain the product of ALP couplings to nucleons and top quarks at the level of $O(0.01$\,TeV$^{-1})$. Notably,  ALP-photon collisions naturally provide the leading probe of ALP flavor-violating couplings to the top quark. We suggest that the 2016 $p$Pb dataset collected at CMS, ATLAS, and LHCb  should be explored for evidence of such collisions.

\end{abstract}

\maketitle

\section{Introduction \label{se:Intro}}

Scalar fields with approximate shift symmetry are ubiquitous in the  landscape of theories beyond the Standard Model (SM). 
They naturally arise from the  compactification of higher-dimensional gauge fields \cite{Dienes:1999gw,Choi:2003wr,Cox:2019rro} and of higher-form  fields from string theory \cite{Conlon:2006tq,Svrcek:2006yi,Arvanitaki:2009fg}. They also arise as  pseudo-Goldstone bosons resulting from the spontaneous breaking of approximate global symmetries \cite{Masso:1995tw}, as happens in the Peccei-Quinn mechanism \cite{Peccei:1977hh}.  In this context, the corresponding particle is called the axion, and it is common to refer to any of these scalar excitations as an \textit{axion-like} particle (ALP).
 Axions and ALPs play an important role in cosmology, such as inflation and the cosmological constant problem, and are excellent candidates for dark matter \cite{Marsh:2015xka}.
In a sense, ALPs provide a versatile portal to new physics, linking open problems in particle physics, cosmology, and fundamental theory.

Axion-like particles are traditionally  searched for in laboratory experiments such as light-shining-through-the-wall setups \cite{Anselm:1985obz,VanBibber:1987rq}, in emission from the Sun (helioscopes) \cite{Sikivie:1983ip} and as dark matter (haloscopes), see \cite{ParticleDataGroup:2024cfk, Giannotti:2024xhx, Ringwald:2024uds} for reviews. Other probes arise from astrophysics \cite{Raffelt:2006cw}, fifth force searches \cite{Ferrer:1998ue,VanTilburg:2024xib,Barbosa:2024pkl,Grossman:2025cov,Cheng:2025fak},  meson properties \cite{Goudzovski:2022vbt,Bauer:2021mvw}, production  at colliders \cite{Brivio:2017ije,Knapen:2016moh,Bauer:2018uxu, dEnterria:2021ljz} --- where limits have been recently reported by the LHC experiments \cite{CMS:2018erd,ATLAS:2020hii,TOTEM:2023ewz, ATLAS:2023zfc, ATLAS:2023ian}, and more.

In this letter, we investigate a phenomenon that  has so far remained unexplored:  high-energy ALP collisions. Where  could ALP collisions  occur in the real world? Consider  photons at the Large Hadron Collider (LHC). Although  the LHC collides protons ($p$) and/or ions ($A$), it can also be viewed  as a photon collider.  This is because, at the semi-classical level, the relativistic beams of hadrons constitute intense electromagnetic sources \cite{Jackson:1998nia, BAUR1990786, Cahn:1990jk}. This strong flux of photons appears at the quantum level via the factorization of amplitudes with forward kinematics \cite{peskin2018introduction}.   
This remarkable fact has led to a forward physics program at the LHC, enabling elegant probes of QED and of physics beyond the SM in ultraperipheral collisions (UPCs) \cite{Baltz:2007kq,Klein:2020fmr}. Notably, the phenomenon of light-by-light scattering \cite{dEnterria:2013zqi} was observed for the first time at ATLAS in PbPb-induced photon collisions \cite{ATLAS:2020hii}. 

Following the same logic, if an ALP exists and couples to hadrons, relativistic beams emit a coherent ALP field.  This emission of a flux of ALPs, sometimes referred to as ``axion bremsstrahlung'', was computed for protons in \cite{BLUMLEIN2014320, Ritz:2021}, and has been used in a number of ways in searches at colliders \cite{Muyuan:2023, Belle:2020, Bauer:2017} and  beam dumps \cite{Tsai:86, liu:2017}. 
However, the phenomena of ALP-ALP and ALP-photon collisions, that follow in analogy to photon-photon collisions, have never been exploited. In this letter, we initiate the investigation of ALP collisions  at the LHC. We study where to best find them, and what ALP-SM couplings are best probed by such processes.  

\section{Axion Couplings}

The  interactions of the  ALP quantum field $a(x^\mu)$ relevant to this study are described at the electroweak scale by the following effective Lagrangian \cite{Bauer:2021mvw}:
\begin{align} \label{eq:L_EW}
 & {\cal L}    = \frac{(\partial_\mu a)^2 }{2}-\frac{m^2_a a^2}{2}   + c_{GG}\frac{\alpha_s}{4\pi}\frac{a}{f} G^{a}_{\mu\nu} \tilde G^{a,\mu\nu}
 \\ & + \frac{\partial^\mu a}{f}\bigg( 
\bar u_L {\bm k}_U \gamma_\mu u_L  + 
\bar u_R {\bm k}_u \gamma_\mu u_R  + 
\bar d_L {\bm k}_D \gamma_\mu d_L  + \ldots
\bigg), \nonumber
\end{align}
where $G^a_{\mu\nu}$ is the gluon field, $u_L$ is a 3-vector encoding the three generations of left-handed quark fields, etc.  
%$\alpha_s$ is the strong coupling constant, 
The ${\bm k}_i$ are Hermitian $3\times 3$ matrices in generation space. The fermion fields are in their mass basis. 
In particular, the $a  f_i\bar f_i$ flavor-conserving couplings take the form $c_{u_iu_i} = [k_u - k_U]_{ii}$.   
% The axion-quark coupling is pseudoscalar if ${\bm k}_U=-{\bm k}_u$, otherwise CP violation can occur. 

Below the QCD scale, the ALP coupling to nucleons $N=(p,n)$ is determined to be 
\begin{align} \label{eq:L_N}
{\cal L}_{\rm nucleons} &=  -i g_{N} a \bar N \gamma_5 N,  \\
g_{p,n} &= \frac{m_N}{2f} \left(g_0 (c_{uu}+c_{dd}+2c_{GG})\pm g_A \Delta c_{ud} \right), \nonumber\\
\Delta c_{ud} & = c_{uu} -c_{dd}+2c_{GG}\frac{m_d-m_u}{m_d+m_u}\,, \nonumber
\end{align}
with $g_0\approx 0.44$, $g_A\approx 1.25$, using chiral perturbation theory techniques, and assuming $m_a\ll m_\pi$. When $m_a$ is closer to the pion mass, ALP-pion mixing can substantially increase  $g_{ap}$ \cite{Bauer:2021mvw}.

\section{Axion Fluxes from Protons and Ions \label{se:Fluxes}}

To describe ALP collisions, we need to know how ALPs are emitted from protons and ions. We focus on the elastic component of the fluxes --- that will be mostly selected by our requirements on final states. These ALP fluxes  are described by collinear approximations in the spirit of the  Weizs\"acker-Williams-type approximations for photons.

From the QFT viewpoint, we derive the ALP flux  from the factorization of the ALP emission cross section in the forward region. 
Using the ALP-nucleon EFT of Eq.~\eqref{eq:L_N}, the ALP and photon fluxes from the proton are given by:
\begin{align}
f_{a|p}(x) &= \frac{g_p^2 }{16 \pi^2} x \log{\left(\frac{E_p^2}{m_p^2 + \frac{1-x}{x^2}  m_a^2} \right)}, \label{eq:WW_axion}
 \\
f_{\gamma|p}(x) &= \frac{\alpha}{2 \pi}\frac{1 + (1-x)^2}{x}\log{\left(\frac{E_p^2}{m_p^2} \right)},\label{eq:WW}
\end{align}
where $E_p$ and $m_p$ are the energy and proton mass, respectively. The result Eq.~\eqref{eq:WW_axion} agrees with the flux obtained by integrating the splitting function for axion bremsstrahlung~\cite{Boiarska:2019jym}.
The difference in $x$-dependence between $f_{a|p}$ and $f_{\gamma|p}$ is tied to the fact that ALP emission requires a helicity flip of the proton while photon emission does not.

For  fluxes of particles from heavy ions, a more accurate description is obtained using the semi-classical approach of virtual quanta~\cite{Jackson:1998nia}. In analogy with the electromagnetic case, 
we treat the ALP field classically.  
Its equation of motion sourced by a relativistic point-like charge is
\begin{equation}
    \square a(t,\vec{x}) = (Z g_p+ (A-Z)g_n) \,\delta^{(3)}(\vec{x}-\vec{x}^\prime(t))\,.
\end{equation}
In the following, we take $g_p\approx g_n$ for simplicity. 
We evaluate the field at a fixed transverse distance $\vec{b}$ from the ion trajectory. The number distribution of equivalent ALPs per unit area is then obtained from the energy density of the field in frequency space:
\begin{equation}
    \frac{d^2 f_{a/A}}{d b^2}(x, \vec{b}) = \frac{A^2 g_p^2}{16 \pi^3 x}(x M_{A})^2\left(K_0^2\left(\xi \right) + \frac{1}{ \gamma^2}K_1^2\left(\xi\right) \right)\,,
\end{equation}
where $\xi = x M_{A}|\vec{b}|$, $M_{A}$, and $\gamma$ are the ion mass and Lorentz factor, and $K_n$ are the modified Bessel functions of the second kind.

The total distribution is obtained by integrating over the possible impact parameters, from the nuclear radius $R$ to infinity. The resulting ALP and photon fluxes from ions are given by:
\begin{align}
    f_{a|A}(x) &= \frac{A^2 g_p^2}{8 \pi^2 \gamma^2 x} y \Big[ K_0\left(y\right) K_1\left(y\right)  \label{eq:edff_axion}\\ & \quad \quad \quad \quad \quad - \frac{y}{2}(1-\gamma^2)\left(K_1^2\left(y\right) - K_0^2\left(y\right)\right) \Big],\nonumber\\
    f_{\gamma|A}(x) &= \frac{2 Z^2 \alpha}{\pi x} y \Big[ K_0\left(y\right)K_1\left(y\right)\label{eq:edff} \\ & \quad \quad \quad \quad \quad \quad\quad \quad -\frac{y}{2} \left(K_1^2\left(y\right) - K_1^2\left(y\right)\right) \Big]\nonumber\,,
\end{align}
where $y = x M_{A} R$. 
The fluxes are exponentially suppressed for high frequencies,   $y \gg 1$. In the low-frequency regime $y \ll 1$, the ALP flux scales as $f_{a|A} \sim \gamma^{-2}  x^{-1} \log{(1/y)}$, while $f_{\gamma|A} \sim x^{-1} \log{(1/y)}$. Since $\gamma\gg1$, the ALP flux from ions is significantly suppressed relative to the photon flux in the ultrarelativistic limit.
From the semiclassical view, this suppression occurs because the scalar field has no transverse degrees of freedom. In contrast, in the electromagnetic  field case, the dominant contribution comes from the transverse component of the electric field, which is amplified by the relativistic boost as $E_\perp^\prime \sim \gamma E_\perp$ leading to the relative $\gamma^2$ factor between $f_{\gamma|A}$ and $f_{a|A}$.

\begin{figure}[t]
    \centering
            \includegraphics[width=0.538\linewidth,trim={0.23cm 0cm 0.1cm 0cm},clip]{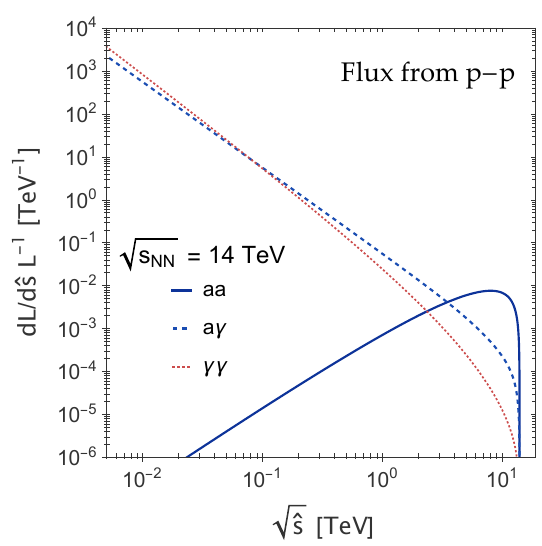}
    \includegraphics[width=0.45\linewidth,trim={1.78cm 0cm 0cm 0cm},clip]{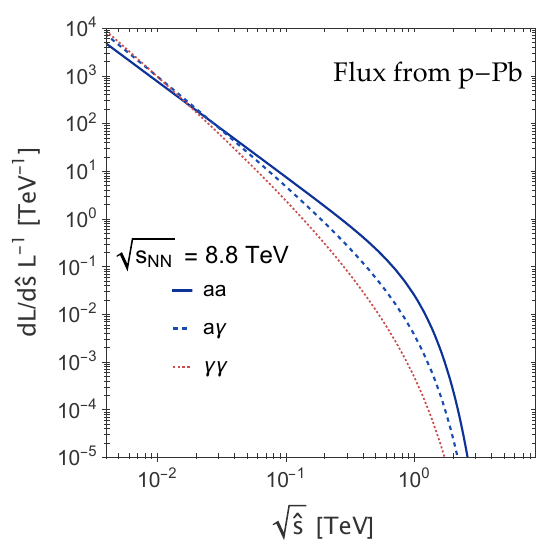}

    \caption{Normalized differential luminosities of the $aa$, $a\gamma$, $\gamma\gamma$ collisions triggered  from the $pp$ and $p$Pb UPCs occurring at the LHC. $\sqrt{s_{NN}}$ is the center-of-mass energy per nucleon.   
    }
    \label{fig:luminosity}
\end{figure}

In the ideal ultraperipheral collision of two nuclei $N$, $N'$ (with $N=p,A$) emitting respectively quasi-real particles $a$, $b$,  the total cross section 
is governed by the joint number distribution of the two quasi-real particles carrying fractions of the total energy of each nucleus $x_a$, $x_b$: $f_{a|N}(x_a)f_{b|N'}(x_b)dx_a dx_b$. 
The total UPC cross section $\sigma_{N,N'}$ for a  sub-process $a+b \rightarrow X$ with cross section $\sigma_{a + b \rightarrow X}$ is 
\begin{equation}
    \sigma_{N,N'} = \int d\hat{s} \frac{dL^{N,N'}_{a,b}}{d\hat{s}}\sigma_{a+b \rightarrow X}(\hat{s})\,,
\end{equation}
where the effective differential luminosity for producing a system with invariant mass $\sqrt{\hat{s}}$ is given by
\begin{align}
    &\frac{dL^{N,N'}_{a,b}}{d\hat{s}} = \label{eq:diff_luminosity} \frac{1}{s} \int_{\hat{s}/s}^1 dx_a dx_b f_{a|N}(x_a) f_{b|N'}(x_b) \delta\left(\frac{\hat{s}}{s} - x_a x_b \right)\,
\end{align}

The normalized profiles of ${d L^{pp}}/{d\hat s}$ and ${d L^{p{\rm Pb}}}/{d\hat s}$ are shown in Fig.\,\ref{fig:luminosity}. The curves show that $aa$ and $a\gamma$ collisions occur typically at center-of-mass energy higher  than the one for $\gamma\gamma$. This is due to the specific structure of the ALP fluxes, stemming from the ALP being scalar.

The validity of the ALP-nucleon EFT imposes that the ALP virtuality $Q$ is bounded from above by the proton mass, i.e. by the inverse proton radius $R_p^{-1}\sim m_p$. Unlike for the photon, the ALP-nucleon form factors that smoothly implement the cutoff on $Q^2$   are unknown experimentally. Still, since these form factors come from the composite structure of the proton, which is governed by QCD,  it is reasonable to assume that the scale of the response of the proton to an ALP probe is similar to a photon probe. We thus use the bound $Q^2 \leq Q_{\rm max}^2 \approx m^2_p$.
 This implies an upper bound on $x$, that is for instance $x\leq 0.62$ when $m_a\ll m_p$.    Notice there is some intrinsic uncertainty associated to this cutoff, for instance  a variation of $Q_{\rm max}$ by $\pm 10\%$ results in a variation of the total cross section  by $\sim 13\%$.

We have interfaced the ALP fluxes with \texttt{MadGraph5}\,\cite{Alwall:2014hca}.

\section{Axion Collisions at the LHC}

Where are ALP collisions  most likely to happen at the LHC? 

Let us first select the collision mode. A key feature of our fluxes is that  ALP emission from ions is suppressed by an extra $\frac{1}{\gamma^2}$ factor. 
Therefore, while the yields for $\gamma\gamma$ from the modes $pp, p{\rm A}, {\rm AA}$ have relative scaling $1$:$Z^2$:$Z^4$, the yields from $a\gamma$ and $aa$ are respectively $1$:$Z^2+\frac{A^2}{\gamma^2}$:$\frac{Z^2 A^2}{\gamma^2}$ and 
$1$:$\frac{A^2}{\gamma^2}$:$\frac{A^4}{\gamma^4}$. 
Since  $\gamma=O(10^3)$ at the LHC,
   we have $Z^2\gg \frac{A^2}{\gamma^2}$, which is enough to conclude  that  the $pA$ mode is the most interesting, with $a$ emitted from the proton and $\gamma$ from the ion. 
    
    Focusing on the production of fermion-antifermion $ \bar f f ^{'} $  pairs, the full ultraperipheral process is:    \includegraphics[width=1.1\linewidth,trim={5cm 6cm 4cm 6.5cm},clip]{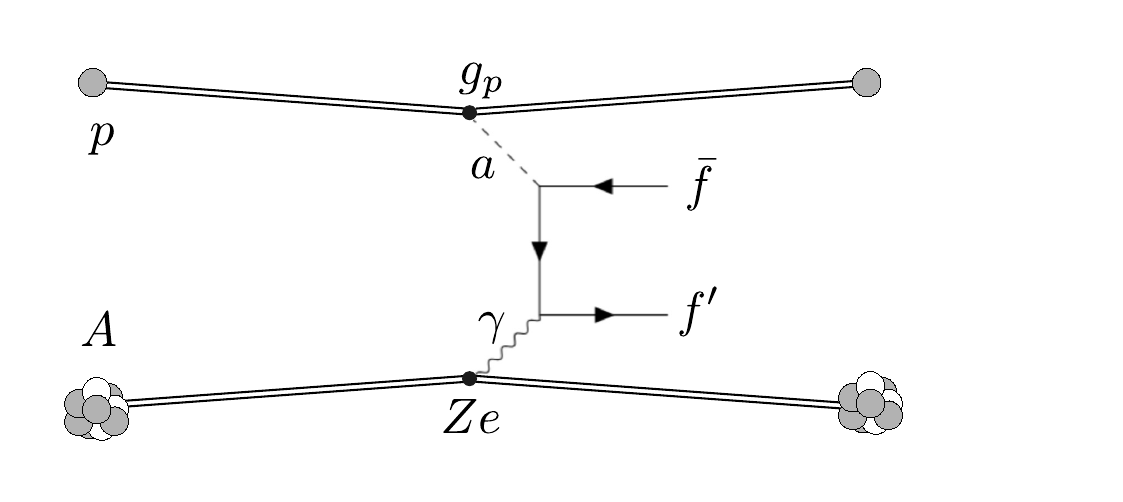}
 Upon integration by parts in Eq.~\eqref{eq:L_EW}, the non-derivative ALP couplings scale with the fermion masses, hence the third generation is the most favorable.  We thus focus on the top quark sector.

The key cross sections to compare the collision modes are as follows: 
\begin{table}[h!]
    \centering
    \begin{tabular}{|c|c|c|c|}
    \hline
  UPC  & $pp$ & $p$O & $p$Pb \\
   \hline
     \hline
    $a\gamma \to t \bar t$  &  $3.2 \cdot 10^{-3}\frac{g^2c^2_{tt}}{f^2}$ pb  & $10.3 \frac{g^2c^2_{tt}}{f^2} $ pb &  $1.4 \cdot 10^{4}\frac{g^2c^2_{tt}}{f^2}$ pb \\
    \hline
        $a\gamma \to t \bar c$  &  $3.4\cdot 10^{-3}\frac{g^2|c_{tc}|^2}{f^2}$ pb  & $ 9.9\frac{g^2|c_{tc}|^2}{f^2} $ pb &  $2.3\cdot 10^4\frac{g^2|c_{tc}|^2}{f^2}$ pb \\
        \hline
        \hline 
   $\gamma\gamma\to t\bar t$  &   $1.8 \cdot 10^{-4}$ pb &  $1.6 \cdot 10^{-3}$ pb & $3.0 \cdot 10^{-2}$ pb \\
       \hline
    \end{tabular}
\end{table} \\
 with the ALP coupling scale $f$ in units of TeV, and $c_{tc}=[k_u-k_U]_{23}$. We assume the nominal c.o.m. energies $\sqrt{s_{NN}}=14, 9.9, 8.8$\,TeV, respec\-tively, and 
 require  $p_T(j)>20$\,GeV for the jets.

We can see that the cross-sections for all processes increase with the ion mass. However, the $a\gamma$ cross-sections  increase much more than the $\gamma\gamma$ ones, an effect related to the larger fraction of energy carried by the ALP. 
% , see Fig.\,\ref{fig:luminosity}. 
In the following, our analyses focus thus on the $p$Pb collisions, which offer both a large signal and the best signal-to-background ratio.

\section{Analysis Framework}

The effective Lagrangian is implemented using Universal FeynRules Output (UFO) format and interfaced with {\tt{Madgraph5}}. Event samples are generated for $\gamma\gamma\to t\bar{t}$, $\gamma a\to t\bar{t}$, and $\gamma a\to tj$ processes in $p$-Pb collisions at the same 8.16~TeV colliding energy as the LHC 2016 data-taking, with final-state jets required to have $p_{T}(j) >$~20~GeV. The ALP flux from the proton is modeled in {\tt{Madgraph5}} following Eq.~\eqref{eq:WW_axion} while the photon flux from Pb is described by the electric dipole form factor implemented in {\tt{gammaUPC}} \cite{Shao:2022cly}, as given in Eq.~\eqref{eq:edff}. A transverse momentum boost is then applied to the event samples, distributed according to $p_T/(p_T^2+Q_0^2)^2$, with $Q_0^2 \approx 0.71$~GeV$^2$ \cite{Budnev:1975poe,daSilveira:2014jla}. The modeling is validated against predictions from {\tt{Superchic}} \cite{Harland-Lang:2020veo}.

\begin{figure}[t]
    \centering
    \includegraphics[width=0.511\linewidth,trim={8.302cm 3.4cm 7.6cm 4.7cm},clip]{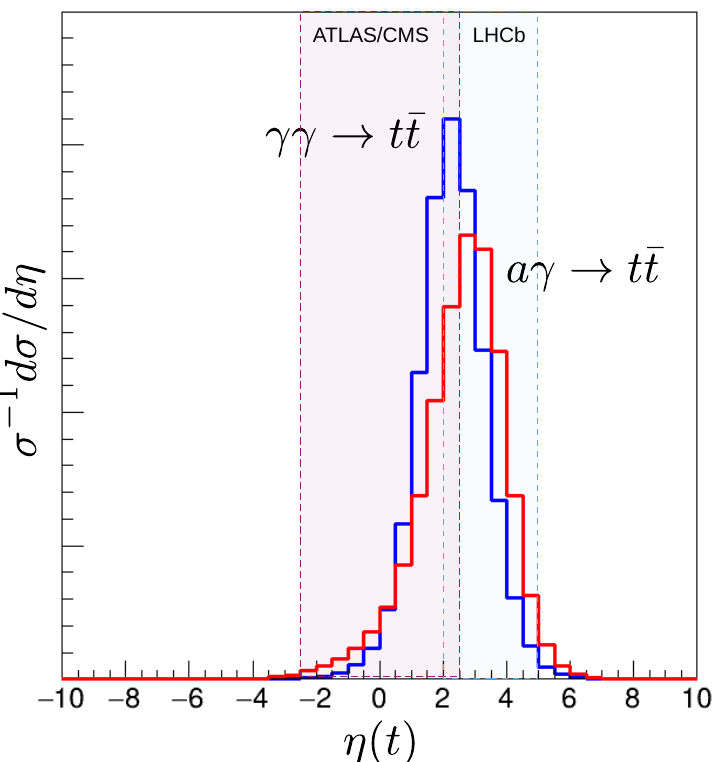}
      \includegraphics[width=0.476\linewidth,trim={9.05cm 3.3cm 7.6cm 4cm},clip]{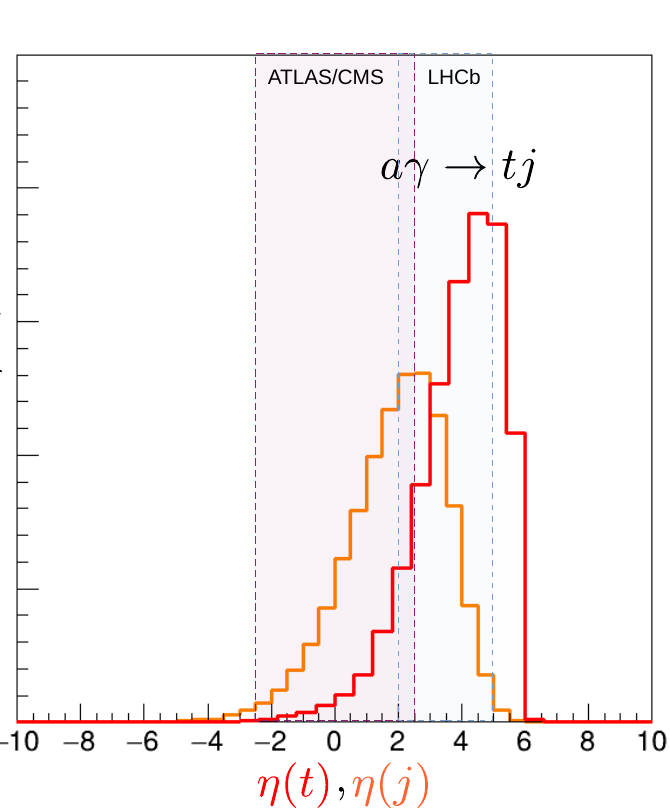}
    \caption{Pseudo-rapidity distributions of the final states from the collisions $a\gamma\to t\bar t $, $\gamma\gamma\to t\bar t $ (left) and $a\gamma\to t j $ (right)  in the $p$Pb mode.   }
    \label{fig:eta}
\end{figure}

For event selection, we focus on the exclusive contribution, as non-exclusive events involving proton dissociation and QCD backgrounds can be suppressed by vetoing calorimeter towers above a certain noise \cite{CMS:2018bbk,CMS:2018nhd}. No specific selection is applied to the final-state top quarks or jets, in order to assess the raw event yield within the acceptances of the central detectors. Proton tagging is not required in this analy\-sis, although it can be exploited in single proton-tagged events --- notably, $p$Pb collisions at the LHC exhibit negligible pileup, typically with $\langle\mu\rangle\sim$~0.1 \cite{CMS:2018fkg}. The pseudorapidity distributions for signal and background are presented in Fig.~\ref{fig:eta}, along with the detector acceptances of LHC experiments. We find that 48\% (21\%) of $a\gamma\to t\bar{t}$ events and 36\% (8\%) of $a\gamma\to t\bar{c}$ events fall within the LHCb (CMS/ATLAS) acceptance.

Tagging/reconstruction efficiencies and systematic uncertainties are not expected to significantly alter the  picture.
Lepton and jet reconstruction are based on particle-flow algorithms, which offer high efficiency in particle identification. For jets, a typical transverse momentum threshold of 20~GeV is sufficient to ensure high trigger efficiency \cite{LHCb:2018usb,LHCb:2020frr,ATLAS:2019guf}. These algorithms also provide excellent performance in reconstructing high-$p_T$ leptons, including those from highly boosted top quarks with $p_T > 400$~GeV \cite{CMS:2022kqg}. When combined with jet tagging techniques, the event selection strategy  enables efficient reconstruction of $t\bar{t}$ events. In particular, tailored clustering procedures for boosted $t\bar{t}$ final state have shown to be very effective \cite{CMS:2022kqg}. Regarding systematic uncertainties, jet energy corrections typically remain $\lesssim$10\% for jets with $p_{T}>$~20~GeV. Furthermore, the misidentification rate of light-flavor jets in $t\bar{t}$ events is about 1\%~\cite{CMS:2015jwh}.

From the above we conclude that   $p$Pb UPCs at the LHC present an opportunity to search for ALP-photon collisions. Notably, a dataset collected in 2016 is already available for analysis.

\section{Projected Sensitivity and Discussion}

Let us  evaluate how sensitive the LHC experiments are to the specific ALP-photon collision process investigated here.

At the typical integrated luminosity $L_{\rm int}$ considered, the number of background events is much lower than $1$. Thus we apply the simple statistical framework of \cite{Baldenegro:2018hng}. 
Based on the small background, our projected data assume that no event is observed, which sets an upper bound on the signal event rate. A $95\%$ credibility level (CL) corresponds to $\sigma\approx 3 L_{\rm int}^{-1}$ \cite{Baldenegro:2018hng}.

Combining this statistical method with the experimental analysis framework described above provides the expected sensitivity to the exclusive $a\gamma \to t\bar t$ and $a\gamma \to tj,\bar tj$ processes. 
This sensitivity  is expressed in terms of upper bounds on the ALP couplings. We focus on the ${\bm k}_u=-{\bm k}_U$ case, for which there is no CP violation. 
 The probed couplings are  $c_{tt}$ and $c_{tq}$, where $c_{tq}=\sqrt{|c_{tu}|^2+|c_{tc}|^2}$. The latter combination appears because the processes  $a\gamma\to t \bar u$, $a\gamma\to t \bar c$ have,  to a very good approximation, the same cross section such that both contribute equally to  single-top+jet final state. 

We consider the $L_{\rm int}$ corresponding   to the 2016 $p$Pb data collected at ATLAS and CMS ($220 $\,nb$^{-1}$ each) and LHCb ($36 $\,nb$^{-1}$), as well as the realistic projection  $1{\rm\,pb}^{-1}$ \cite{dEnterria:2025jgm}. 
Putting everything together, we obtain
the  $95$\%CL bounds
\,\footnote{We mention that for ${\bm k}_u\neq-{\bm k}_U$, CP violation can occur such that the $t$ and $\bar t$ distributions  differ. The numbers in the table can  change by a $O(1)$ factor, and an asymmetry arises between  the $tj$, $\bar t j$ yields. }
\begin{table}[h!]
    \centering
    \begin{tabular}{|c|c|c|c|c|}
    \hline
        Det. &  \multicolumn{2}{c}{ATLAS/CMS}  & \multicolumn{2}{|c|}{ LHCb}   \\
        \hline
        $L_{\rm int}$ & 220 nb$^{-1}$  & 1 pb$^{-1}$ & 36 nb$^{-1}$  & 1 pb$^{-1}$ \\
         \hline
      $\frac{g c_{tt}}{f}$   & $(7.2$\,TeV)$^{-1}$ &  $(15.3$\,TeV)$^{-1}$ &  $(4.4$\,TeV)$^{-1}$ &   $(23.1$\,TeV)$^{-1}$ \\
      \hline
      $\frac{g c_{tq}}{f}$   & $(11.5$\,TeV)$^{-1}$ &  $(24.5$\,TeV)$^{-1}$ &  $(9.9$\,TeV)$^{-1}$ &   $(52.0$\,TeV)$^{-1}$ \\
      \hline      
    \end{tabular}
\end{table}  \\

Let us discuss how these projected  bounds compare to the existing bounds on ALP couplings.

A combination of the ALP couplings $c_{uu},c_{dd},c_{GG}$ inside the ALP-proton coupling given in Eq.~\eqref{eq:L_N} is constrained by kaon decays \cite{MartinCamalich:2020dfe,Goudzovski:2022vbt}. Specifically, the process $K^+\to\pi^++ a$  leads to anomalous branching ratios that are constrained by the NA62 experiment \cite{NA62:2021zjw,NA62:2020xlg}. The exclusion regions are  however complex and mass-dependent. For example, if the $c_{uu,dd}$ dominate,  the kaon bounds vanish completely for $m_a\geq 350$\,MeV. 
Hence we  treat the $g$ coupling as an independent parameter. 

The $c_{tt}$ coupling is constrained individually in the $m_a\lesssim 5$\,GeV region by the $B$-decays \cite{Esser:2023fdo}. Specifically, the  branching ratio Br$(B\to K +{\rm inv})$ is constrained by BaBar to $\frac{|c_{tt}|}{f}\lesssim 1.5 \cdot 10^{-3}$~TeV$^{-1}$ \cite{BaBar:2013npw}. Notice that $|c_{tt}|$  vanishes if $[k_U]_{33}\approx [k_u]_{33}$.

A strategy to bound the $c_{tc}$, $c_{tu}$ couplings at the LHC using displaced vertices has been proposed in \cite{Cheung:2024qve}, but applies to ALP masses  of order $O(10)$\,GeV.

Comparing these bounds to our results, we conclude that the projected sensitivity to $c_{tt}$ from the ALP-photon collision $a\gamma\to t\bar t$ is weaker than the one from $B$-decays. 
On the other hand, the $c_{tc}$, $c_{tu}$ couplings are so far unconstrained in the sub-GeV mass range probed in our exclusive processes. We thus find  that ALP-photon collisions at the LHC provide a leading probe of the $c_{tc}$ and $c_{tu}$ couplings.

\section{Summary}

In this letter, we  point out that if a sub-GeV ALP couples to nucleons, then ALP-ALP and ALP-photon collisions can arise from ultraperipheral collisions at the LHC. 
 This phenomenon provides a new probe of  ALPs at TeV-scale  energies. 
 Due to the strong suppression of ALP fluxes from  ions, the most relevant collision mode is the proton-ion one, particularly $p$Pb.  
Focusing on the top quark sector,
we demonstrate that ALP-photon collisions from $p$Pb UPCs produce a sizable amount of top pairs and top$+$jet final states, while having vanishing Standard Model background. The latter channel provides a clean probe of  the ALP-top flavor-violating couplings, which remain so far unconstrained for sub-GeV ALPs. 
These ALP-photon collisions can be readi\-ly searched for in the existing 2016 $p$Pb  datasets collected at  {ATLAS}, CMS and LHCb.

\section*{acknowledgments}

% We thank     César Bernardes and David d'Enterria for useful discussions. 
% The work of MC and SF was supported in part by the S\~ao Paulo Research Foundation (FAPESP), under respectively grants 2022/15878-0 and 2021/10128-0.  GGS acknowledges funding from FAPERGS and the Brazilian agency Conselho Nacional de Desenvolvimento Científico e Tecnológico (CNPq) with grant CNPq/315246/2023-5.

We thank César Bernardes and
David d’Enterria for useful discussions. The work of
S. B., M. C., and S. F. is supported by the São Paulo
Research Foundation (FAPESP), under respectively
Grants No. 2025/05571-3, No. 2022/15878-0, and
No. 2021/10128-0. The work of S. B. has been supported
in part by Grant No. 001 of CAPES. G. G. d. S. acknowledges funding from FAPERGS and the Brazilian agency
Conselho Nacional de Desenvolvimento Científico e
Tecnológico (CNPq) with Grant No. CNPq/315246/2023-5.

\bibliographystyle{JHEP}
\bibliography{referencias}

\providecommand{\href}[2]{#2}\begingroup\raggedright\begin{thebibliography}{10}

\bibitem{Dienes:1999gw}
K.~R. Dienes, E.~Dudas, and T.~Gherghetta, {\it {Invisible axions and large
  radius compactifications}},  {\em Phys. Rev. D} {\bf 62} (2000) 105023,
  [\href{http://arxiv.org/abs/hep-ph/9912455}{{\tt hep-ph/9912455}}].

\bibitem{Choi:2003wr}
K.-w. Choi, {\it {A QCD axion from higher dimensional gauge field}},  {\em
  Phys. Rev. Lett.} {\bf 92} (2004) 101602,
  [\href{http://arxiv.org/abs/hep-ph/0308024}{{\tt hep-ph/0308024}}].

\bibitem{Cox:2019rro}
P.~Cox, T.~Gherghetta, and M.~D. Nguyen, {\it {A Holographic Perspective on the
  Axion Quality Problem}},  {\em JHEP} {\bf 01} (2020) 188,
  [\href{http://arxiv.org/abs/1911.09385}{{\tt arXiv:1911.09385}}].

\bibitem{Conlon:2006tq}
J.~P. Conlon, {\it {The QCD axion and moduli stabilisation}},  {\em JHEP} {\bf
  05} (2006) 078, [\href{http://arxiv.org/abs/hep-th/0602233}{{\tt
  hep-th/0602233}}].

\bibitem{Svrcek:2006yi}
P.~Svrcek and E.~Witten, {\it {Axions In String Theory}},  {\em JHEP} {\bf 06}
  (2006) 051, [\href{http://arxiv.org/abs/hep-th/0605206}{{\tt
  hep-th/0605206}}].

\bibitem{Arvanitaki:2009fg}
A.~Arvanitaki, S.~Dimopoulos, S.~Dubovsky, N.~Kaloper, and J.~March-Russell,
  {\it {String Axiverse}},  {\em Phys. Rev. D} {\bf 81} (2010) 123530,
  [\href{http://arxiv.org/abs/0905.4720}{{\tt arXiv:0905.4720}}].

\bibitem{Masso:1995tw}
E.~Masso and R.~Toldra, {\it {On a light spinless particle coupled to
  photons}},  {\em Phys. Rev. D} {\bf 52} (1995) 1755--1763,
  [\href{http://arxiv.org/abs/hep-ph/9503293}{{\tt hep-ph/9503293}}].

\bibitem{Peccei:1977hh}
R.~D. Peccei and H.~R. Quinn, {\it {CP Conservation in the Presence of
  Instantons}},  {\em Phys. Rev. Lett.} {\bf 38} (1977) 1440--1443.

\bibitem{Marsh:2015xka}
D.~J.~E. Marsh, {\it {Axion Cosmology}},  {\em Phys. Rept.} {\bf 643} (2016)
  1--79, [\href{http://arxiv.org/abs/1510.07633}{{\tt arXiv:1510.07633}}].

\bibitem{Anselm:1985obz}
A.~A. Anselm, {\it {Arion $\leftrightarrow$ Photon Oscillations in a Steady
  Magnetic Field. (In Russian)}},  {\em Yad. Fiz.} {\bf 42} (1985) 1480--1483.

\bibitem{VanBibber:1987rq}
K.~Van~Bibber, N.~R. Dagdeviren, S.~E. Koonin, A.~Kerman, and H.~N. Nelson,
  {\it {Proposed experiment to produce and detect light pseudoscalars}},  {\em
  Phys. Rev. Lett.} {\bf 59} (1987) 759--762.

\bibitem{Sikivie:1983ip}
P.~Sikivie, {\it {Experimental Tests of the Invisible Axion}},  {\em Phys. Rev.
  Lett.} {\bf 51} (1983) 1415--1417. [Erratum: Phys.Rev.Lett. 52, 695 (1984)].

\bibitem{ParticleDataGroup:2024cfk}
{\bf Particle Data Group} Collaboration, S.~Navas et~al., {\it {Review of
  particle physics}},  {\em Phys. Rev. D} {\bf 110} (2024), no.~3 030001.

\bibitem{Giannotti:2024xhx}
M.~Giannotti, {\it {Status and Perspectives on Axion Searches}},  12, 2024.
\newblock \href{http://arxiv.org/abs/2412.08733}{{\tt arXiv:2412.08733}}.

\bibitem{Ringwald:2024uds}
A.~Ringwald, {\it {Review on Axions}},  4, 2024.
\newblock \href{http://arxiv.org/abs/2404.09036}{{\tt arXiv:2404.09036}}.

\bibitem{Raffelt:2006cw}
G.~G. Raffelt, {\it {Astrophysical axion bounds}},  {\em Lect. Notes Phys.}
  {\bf 741} (2008) 51--71, [\href{http://arxiv.org/abs/hep-ph/0611350}{{\tt
  hep-ph/0611350}}].

\bibitem{Ferrer:1998ue}
F.~Ferrer and J.~A. Grifols, {\it {Long range forces from pseudoscalar
  exchange}},  {\em Phys. Rev. D} {\bf 58} (1998) 096006,
  [\href{http://arxiv.org/abs/hep-ph/9805477}{{\tt hep-ph/9805477}}].

\bibitem{VanTilburg:2024xib}
K.~Van~Tilburg, {\it {Wake forces in a background of quadratically coupled
  mediators}},  {\em Phys. Rev. D} {\bf 109} (2024), no.~9 096036,
  [\href{http://arxiv.org/abs/2401.08745}{{\tt arXiv:2401.08745}}].

\bibitem{Barbosa:2024pkl}
S.~Barbosa and S.~Fichet, {\it {Background-induced forces from dark relics}},
  {\em JHEP} {\bf 01} (2025) 021, [\href{http://arxiv.org/abs/2403.13894}{{\tt
  arXiv:2403.13894}}].

\bibitem{Grossman:2025cov}
Y.~Grossman, B.~Yu, and S.~Zhou, {\it {Axion forces in axion backgrounds}},
  \href{http://arxiv.org/abs/2504.00104}{{\tt arXiv:2504.00104}}.

\bibitem{Cheng:2025fak}
Y.~Cheng and S.~Ge, {\it {Background-Enhanced Axion Force by Axion Dark
  Matter}},  \href{http://arxiv.org/abs/2504.02702}{{\tt arXiv:2504.02702}}.

\bibitem{Goudzovski:2022vbt}
E.~Goudzovski et~al., {\it {New physics searches at kaon and hyperon
  factories}},  {\em Rept. Prog. Phys.} {\bf 86} (2023), no.~1 016201,
  [\href{http://arxiv.org/abs/2201.07805}{{\tt arXiv:2201.07805}}].

\bibitem{Bauer:2021mvw}
M.~Bauer, M.~Neubert, S.~Renner, M.~Schnubel, and A.~Thamm, {\it {Flavor probes
  of axion-like particles}},  {\em JHEP} {\bf 09} (2022) 056,
  [\href{http://arxiv.org/abs/2110.10698}{{\tt arXiv:2110.10698}}].

\bibitem{Brivio:2017ije}
I.~Brivio, M.~B. Gavela, L.~Merlo, K.~Mimasu, J.~M. No, R.~del Rey, and
  V.~Sanz, {\it {ALPs Effective Field Theory and Collider Signatures}},  {\em
  Eur. Phys. J. C} {\bf 77} (2017), no.~8 572,
  [\href{http://arxiv.org/abs/1701.05379}{{\tt arXiv:1701.05379}}].

\bibitem{Knapen:2016moh}
S.~Knapen, T.~Lin, H.~K. Lou, and T.~Melia, {\it {Searching for Axionlike
  Particles with Ultraperipheral Heavy-Ion Collisions}},  {\em Phys. Rev.
  Lett.} {\bf 118} (2017), no.~17 171801,
  [\href{http://arxiv.org/abs/1607.06083}{{\tt arXiv:1607.06083}}].

\bibitem{Bauer:2018uxu}
M.~Bauer, M.~Heiles, M.~Neubert, and A.~Thamm, {\it {Axion-Like Particles at
  Future Colliders}},  {\em Eur. Phys. J. C} {\bf 79} (2019), no.~1 74,
  [\href{http://arxiv.org/abs/1808.10323}{{\tt arXiv:1808.10323}}].

\bibitem{dEnterria:2021ljz}
D.~d'Enterria, {\it {Collider constraints on axion-like particles}},  in {\em
  {Workshop on Feebly Interacting Particles}}, 2, 2021.
\newblock \href{http://arxiv.org/abs/2102.08971}{{\tt arXiv:2102.08971}}.

\bibitem{CMS:2018erd}
{\bf CMS} Collaboration, A.~M. Sirunyan et~al., {\it {Evidence for
  light-by-light scattering and searches for axion-like particles in
  ultraperipheral PbPb collisions at $\sqrt{s_\mathrm{NN}} =$ 5.02 TeV}},  {\em
  Phys. Lett. B} {\bf 797} (2019) 134826,
  [\href{http://arxiv.org/abs/1810.04602}{{\tt arXiv:1810.04602}}].

\bibitem{ATLAS:2020hii}
{\bf ATLAS} Collaboration, G.~Aad et~al., {\it {Measurement of light-by-light
  scattering and search for axion-like particles with 2.2 nb$^{-1}$ of Pb+Pb
  data with the ATLAS detector}},  {\em JHEP} {\bf 03} (2021) 243,
  [\href{http://arxiv.org/abs/2008.05355}{{\tt arXiv:2008.05355}}]. [Erratum:
  JHEP 11, 050 (2021)].

\bibitem{TOTEM:2023ewz}
{\bf TOTEM, CMS} Collaboration, A.~Tumasyan et~al., {\it {Search for high-mass
  exclusive diphoton production with tagged protons in proton-proton collisions
  at s=13{\,}{\,}TeV}},  {\em Phys. Rev. D} {\bf 110} (2024), no.~1 012010,
  [\href{http://arxiv.org/abs/2311.02725}{{\tt arXiv:2311.02725}}].

\bibitem{ATLAS:2023zfc}
{\bf ATLAS} Collaboration, G.~Aad et~al., {\it {Search for an axion-like
  particle with forward proton scattering in association with photon pairs at
  ATLAS}},  {\em JHEP} {\bf 07} (2023) 234,
  [\href{http://arxiv.org/abs/2304.10953}{{\tt arXiv:2304.10953}}].

\bibitem{ATLAS:2023ian}
{\bf ATLAS} Collaboration, G.~Aad et~al., {\it {Search for short- and
  long-lived axion-like particles in $H\rightarrow a a \rightarrow 4\gamma $
  decays with the ATLAS experiment at the LHC}},  {\em Eur. Phys. J. C} {\bf
  84} (2024), no.~7 742, [\href{http://arxiv.org/abs/2312.03306}{{\tt
  arXiv:2312.03306}}].

\bibitem{Jackson:1998nia}
J.~D. Jackson, {\em {Classical Electrodynamics}}.
\newblock Wiley, 1998.

\bibitem{BAUR1990786}
G.~Baur and L.~Filho, {\it Coherent particle production at relativistic
  heavy-ion colliders including strong absorption effects},  {\em Nuclear
  Physics A} {\bf 518} (1990), no.~4 786--800.

\bibitem{Cahn:1990jk}
R.~N. Cahn and J.~D. Jackson, {\it {Realistic equivalent photon yields in heavy
  ion collisions}},  {\em Phys. Rev. D} {\bf 42} (1990) 3690--3695.

\bibitem{peskin2018introduction}
M.~Peskin, {\em An introduction to quantum field theory}.
\newblock CRC press, 2018.

\bibitem{Baltz:2007kq}
A.~J. Baltz et~al., {\it {The Physics of Ultraperipheral Collisions at the
  LHC}},  {\em Phys. Rept.} {\bf 458} (2008) 1--171,
  [\href{http://arxiv.org/abs/0706.3356}{{\tt arXiv:0706.3356}}].

\bibitem{Klein:2020fmr}
S.~Klein and P.~Steinberg, {\it {Photonuclear and Two-photon Interactions at
  High-Energy Nuclear Colliders}},  {\em Ann. Rev. Nucl. Part. Sci.} {\bf 70}
  (2020) 323--354, [\href{http://arxiv.org/abs/2005.01872}{{\tt
  arXiv:2005.01872}}].

\bibitem{dEnterria:2013zqi}
D.~d'Enterria and G.~G. da~Silveira, {\it {Observing light-by-light scattering
  at the Large Hadron Collider}},  {\em Phys. Rev. Lett.} {\bf 111} (2013)
  080405, [\href{http://arxiv.org/abs/1305.7142}{{\tt arXiv:1305.7142}}].
  [Erratum: Phys.Rev.Lett. 116, 129901 (2016)].

\bibitem{BLUMLEIN2014320}
J.~Blümlein and J.~Brunner, {\it New exclusion limits on dark gauge forces
  from proton bremsstrahlung in beam-dump data},  {\em Physics Letters B} {\bf
  731} (2014) 320--326.

\bibitem{Ritz:2021}
S.~Foroughi-Abari and A.~Ritz, {\it Dark sector production via proton
  bremsstrahlung},  {\em Phys. Rev. D} {\bf 105} (May, 2022) 095045.

\bibitem{Muyuan:2023}
J.~Liu, Y.~Luo, and M.~Song, {\it {Investigation of the concurrent effects of
  ALP-photon and ALP-electron couplings in Collider and Beam Dump Searches}},
  {\em J. High Energ. Phys.} {\bf 2023} (2023), no.~104
  [\href{http://arxiv.org/abs/2304.05435}{{\tt arXiv:2304.05435}}].

\bibitem{Belle:2020}
{\bf Belle II} Collaboration, {\it Search for axionlike particles produced in
  ${e}^{+}{e}^{-}$ collisions at belle ii},  {\em Phys. Rev. Lett.} {\bf 125}
  (Oct, 2020) 161806.

\bibitem{Bauer:2017}
M.~Bauer, M.~Neubert, and A.~Thamm, {\it {Collider Probes of Axion-Like
  Particles}},  {\em J. High Energ. Phys.} {\bf 2017} (2017), no.~44
  [\href{http://arxiv.org/abs/1708.00443}{{\tt arXiv:1708.00443}}].

\bibitem{Tsai:86}
Y.~S. Tsai, {\it Axion bremsstrahlung by an electron beam},  {\em Phys. Rev. D}
  {\bf 34} (Sep, 1986) 1326--1331.

\bibitem{liu:2017}
Y.-S. Liu and G.~A. Miller, {\it Validity of the weizs\"acker-williams
  approximation and the analysis of beam dump experiments: Production of an
  axion, a dark photon, or a new axial-vector boson},  {\em Phys. Rev. D} {\bf
  96} (Jul, 2017) 016004.

\bibitem{Boiarska:2019jym}
I.~Boiarska, K.~Bondarenko, A.~Boyarsky, V.~Gorkavenko, M.~Ovchynnikov, and
  A.~Sokolenko, {\it {Phenomenology of GeV-scale scalar portal}},  {\em JHEP}
  {\bf 11} (2019) 162, [\href{http://arxiv.org/abs/1904.10447}{{\tt
  arXiv:1904.10447}}].

\bibitem{Alwall:2014hca}
J.~Alwall, R.~Frederix, S.~Frixione, V.~Hirschi, F.~Maltoni, O.~Mattelaer,
  H.~S. Shao, T.~Stelzer, P.~Torrielli, and M.~Zaro, {\it {The automated
  computation of tree-level and next-to-leading order differential cross
  sections, and their matching to parton shower simulations}},  {\em JHEP} {\bf
  07} (2014) 079, [\href{http://arxiv.org/abs/1405.0301}{{\tt
  arXiv:1405.0301}}].

\bibitem{Shao:2022cly}
H.-S. Shao and D.~d'Enterria, {\it {gamma-UPC: automated generation of
  exclusive photon-photon processes in ultraperipheral proton and nuclear
  collisions with varying form factors}},  {\em JHEP} {\bf 09} (2022) 248,
  [\href{http://arxiv.org/abs/2207.03012}{{\tt arXiv:2207.03012}}].

\bibitem{Budnev:1975poe}
V.~M. Budnev, I.~F. Ginzburg, G.~V. Meledin, and V.~G. Serbo, {\it {The Two
  photon particle production mechanism. Physical problems. Applications.
  Equivalent photon approximation}},  {\em Phys. Rept.} {\bf 15} (1975)
  181--281.

\bibitem{daSilveira:2014jla}
G.~G. da~Silveira, L.~Forthomme, K.~Piotrzkowski, W.~Sch\"afer, and
  A.~Szczurek, {\it {Central $\mu^{+}\mu^{-}$ production via photon-photon
  fusion in proton-proton collisions with proton dissociation}},  {\em JHEP}
  {\bf 02} (2015) 159, [\href{http://arxiv.org/abs/1409.1541}{{\tt
  arXiv:1409.1541}}].

\bibitem{Harland-Lang:2020veo}
L.~A. Harland-Lang, M.~Tasevsky, V.~A. Khoze, and M.~G. Ryskin, {\it {A new
  approach to modelling elastic and inelastic photon-initiated production at
  the LHC: SuperChic 4}},  {\em Eur. Phys. J. C} {\bf 80} (2020), no.~10 925,
  [\href{http://arxiv.org/abs/2007.12704}{{\tt arXiv:2007.12704}}].

\bibitem{CMS:2018bbk}
{\bf CMS} Collaboration, A.~M. Sirunyan et~al., {\it {Measurement of exclusive
  $\Upsilon$ photoproduction from protons in pPb collisions at
  $\sqrt{s_\mathrm{NN}} =$ 5.02 TeV}},  {\em Eur. Phys. J. C} {\bf 79} (2019),
  no.~3 277, [\href{http://arxiv.org/abs/1809.11080}{{\tt arXiv:1809.11080}}].
  [Erratum: Eur.Phys.J.C 82, 343 (2022)].

\bibitem{CMS:2018nhd}
{\bf CMS} Collaboration, A.~M. Sirunyan et~al., {\it {Measurement of charged
  particle spectra in minimum-bias events from proton\textendash{}proton
  collisions at $\sqrt{s}=13\,\text {TeV} $}},  {\em Eur. Phys. J. C} {\bf 78}
  (2018), no.~9 697, [\href{http://arxiv.org/abs/1806.11245}{{\tt
  arXiv:1806.11245}}].

\bibitem{CMS:2018fkg}
{\bf CMS} Collaboration, {\it {CMS luminosity measurement using 2016
  proton-nucleus collisions at nucleon-nucleon center-of-mass energy of 8.16
  TeV}}, .

\bibitem{LHCb:2018usb}
{\bf LHCb} Collaboration, R.~Aaij et~al., {\it {Measurement of forward top pair
  production in the dilepton channel in $pp$ collisions at $\sqrt{s}=13$ TeV}},
   {\em JHEP} {\bf 08} (2018) 174, [\href{http://arxiv.org/abs/1803.05188}{{\tt
  arXiv:1803.05188}}].

\bibitem{LHCb:2020frr}
{\bf LHCb} Collaboration, R.~Aaij et~al., {\it {Measurement of differential $
  b\overline{b} $- and $ c\overline{c} $-dijet cross-sections in the forward
  region of $pp$ collisions at $ \sqrt{s} $ = 13 TeV}},  {\em JHEP} {\bf 02}
  (2021) 023, [\href{http://arxiv.org/abs/2010.09437}{{\tt arXiv:2010.09437}}].

\bibitem{ATLAS:2019guf}
{\bf ATLAS} Collaboration, G.~Aad et~al., {\it {Measurement of the top-quark
  mass in $t\bar{t}+1$-jet events collected with the ATLAS detector in $pp$
  collisions at $\sqrt{s}=8$ TeV}},  {\em JHEP} {\bf 11} (2019) 150,
  [\href{http://arxiv.org/abs/1905.02302}{{\tt arXiv:1905.02302}}].

\bibitem{CMS:2022kqg}
{\bf CMS} Collaboration, A.~Tumasyan et~al., {\it {Measurement of the
  differential $\hbox {t}\overline{\hbox {t}}$ production cross section as a
  function of the jet mass and extraction of the top quark mass in hadronic
  decays of boosted top quarks}},  {\em Eur. Phys. J. C} {\bf 83} (2023), no.~7
  560, [\href{http://arxiv.org/abs/2211.01456}{{\tt arXiv:2211.01456}}].

\bibitem{CMS:2015jwh}
{\bf CMS} Collaboration, V.~Khachatryan et~al., {\it {Search for vector-like T
  quarks decaying to top quarks and Higgs bosons in the all-hadronic channel
  using jet substructure}},  {\em JHEP} {\bf 06} (2015) 080,
  [\href{http://arxiv.org/abs/1503.01952}{{\tt arXiv:1503.01952}}].

\bibitem{Baldenegro:2018hng}
C.~Baldenegro, S.~Fichet, G.~von Gersdorff, and C.~Royon, {\it {Searching for
  axion-like particles with proton tagging at the LHC}},  {\em JHEP} {\bf 06}
  (2018) 131, [\href{http://arxiv.org/abs/1803.10835}{{\tt arXiv:1803.10835}}].

\bibitem{dEnterria:2025jgm}
D.~d'Enterria et~al., {\it {Physics with high-luminosity proton-nucleus
  collisions at the LHC}},  \href{http://arxiv.org/abs/2504.04268}{{\tt
  arXiv:2504.04268}}.

\bibitem{Note1}
We mention that for ${\protect \bm {k}}_u\protect \neq -{\protect \bm {k}}_U$,
  CP violation can occur such that the $t$ and $\protect \bar t$ distributions
  differ. The numbers in the table can change by a $O(1)$ factor, and an
  asymmetry arises between the $tj$, $\protect \bar t j$ yields.

\bibitem{MartinCamalich:2020dfe}
J.~Martin~Camalich, M.~Pospelov, P.~N.~H. Vuong, R.~Ziegler, and J.~Zupan, {\it
  {Quark Flavor Phenomenology of the QCD Axion}},  {\em Phys. Rev. D} {\bf 102}
  (2020), no.~1 015023, [\href{http://arxiv.org/abs/2002.04623}{{\tt
  arXiv:2002.04623}}].

\bibitem{NA62:2021zjw}
{\bf NA62} Collaboration, E.~Cortina~Gil et~al., {\it {Measurement of the very
  rare K$^{+}$\textrightarrow{}$ {\pi}^{+}\nu \overline{\nu} $ decay}},  {\em
  JHEP} {\bf 06} (2021) 093, [\href{http://arxiv.org/abs/2103.15389}{{\tt
  arXiv:2103.15389}}].

\bibitem{NA62:2020xlg}
{\bf NA62} Collaboration, E.~Cortina~Gil et~al., {\it {Search for a feebly
  interacting particle $X$ in the decay $K^{+}\rightarrow\pi^{+}X$}},  {\em
  JHEP} {\bf 03} (2021) 058, [\href{http://arxiv.org/abs/2011.11329}{{\tt
  arXiv:2011.11329}}].

\bibitem{Esser:2023fdo}
F.~Esser, M.~Madigan, V.~Sanz, and M.~Ubiali, {\it {On the coupling of
  axion-like particles to the top quark}},  {\em JHEP} {\bf 09} (2023) 063,
  [\href{http://arxiv.org/abs/2303.17634}{{\tt arXiv:2303.17634}}].

\bibitem{BaBar:2013npw}
{\bf BaBar} Collaboration, J.~P. Lees et~al., {\it {Search for $B \to K^{(*)}
  \nu \overline \nu$ and invisible quarkonium decays}},  {\em Phys. Rev. D}
  {\bf 87} (2013), no.~11 112005, [\href{http://arxiv.org/abs/1303.7465}{{\tt
  arXiv:1303.7465}}].

\bibitem{Cheung:2024qve}
K.~Cheung, F.-T. Chung, G.~Cottin, and Z.~S. Wang, {\it {Quark flavor violation
  and axion-like particles from top-quark decays at the LHC}},  {\em JHEP} {\bf
  07} (2024) 209, [\href{http://arxiv.org/abs/2404.06126}{{\tt
  arXiv:2404.06126}}].

\end{thebibliography}\endgroup

\end{document}